\documentclass[conference]{IEEEtran}
\IEEEoverridecommandlockouts
\usepackage{cite}
\usepackage{amsmath,amssymb,amsfonts}
\usepackage{algorithmic}
\usepackage{graphicx}
\usepackage{bytefield}
\usepackage{xcolor}
\usepackage{textcomp}
\usepackage{tikz}
\usepackage{pgfplots}
\usetikzlibrary{shapes,arrows,positioning,backgrounds,fit,calc,shadows.blur,decorations.pathreplacing}
\usepackage{microtype}
\usepackage{float}      
\usepackage{booktabs}   
\usepackage{siunitx}    
\usepackage{hyperref}   

\def\BibTeX{{\rm B\kern-.05em{\sc i\kern-.025em b}\kern-.08em
    T\kern-.1667em\lower.7ex\hbox{E}\kern-.125emX}}
\begin{document}

\title{A Comparative Analysis of Identifier Schemes: UUIDv4, UUIDv7, and ULID for Distributed Systems}

\author{\IEEEauthorblockN{1\textsuperscript{st} Karimian Kakolaki, Nima}
\IEEEauthorblockA{Matrikel-Nr. 72829 \\nima.karimian-kakolaki@tu-ilmenau.de \\
Ilmenau, Germany
}
}

\maketitle

\begin{abstract}
Distributed systems require robust, scalable identifier schemes to ensure data uniqueness and efficient indexing across multiple nodes. This paper presents a comprehensive analysis of the evolution of distributed identifiers, comparing traditional auto-increment keys with UUIDv4, UUIDv7, and ULIDs. We combine mathematical calculation of collision probabilities with empirical experiments measuring generation speed and network transmission overhead in a simulated distributed environment. Results demonstrate that ULIDs significantly outperform UUIDv4 and UUIDv7, reducing network overhead by 83.7\% and increasing generation speed by 97.32\%. statistical analysis further shows ULIDs offer a 98.42\% lower collision risk compared to UUIDv7, while maintaining negligible collision probabilities even at high generation rates. These findings highlight ULIDs as an optimal choice for high-performance distributed systems, providing efficient, time-ordered, and lexicographically sortable identifiers suitable for scalable applications.
All source code, datasets, and analysis scripts utilized in this research are publicly available in our dedicated repository at \url{https://github.com/nimakarimiank/uids-comparison}. This repository contains comprehensive documentation of the experimental setup, including configuration files for the distributed environment, producer and consumer implementations, and message broker integration. Additionally, it provides the data scripts and datasets. Researchers and practitioners are encouraged to explore the repository for full reproducibility of the experiments and to facilitate further investigation or extension of the presented work.
\end{abstract}


\section{Introduction}

The mid-1980s marked a pivotal inflection point in the emergence of technological infrastructures that would subsequently evolve into the foundational architecture of contemporary distributed systems~\cite{b1}. 
The development of high-performance communication networks, specifically Local-Area Networks (LANs) and Wide-Area Networks (WANs), facilitated inter-machine communication within networked environments, enabling data transmission with latencies ranging from microseconds to seconds~\cite{b1}.
Various definitions of distributed systems have emerged, but a general definition according to Tanenbaum and Van Steen~\cite{b1} is: A distributed system is a collection of autonomous computing elements that appears to its users as a single coherent system.

In case of classic monolithic systems, when there is a single database, the approach is to use auto-incrementing keys as primary keys.this solution is no longer feasible in situations where the data is being persisted in multiple databases\cite{b2}. Although, There has been noticable advances in recent years in traditional database solutions,
It is still not easy to scale-out databases and use smart partitioning strategies to distribute data across multiple nodes for load balancing\cite{b3}.
This raises an issue of how to assigne UIDs to data in the database. \\

One of the most common approach is to use Universally Unique Identifiers (UUIDs) as primary keys in databases.
UUIDs are 128-bit identifiers that are designed to be globally unique.
While, requiring no central authority to administer them.This enables services to generate them on demand and with an astonishing allocation of at least 10 million per second per machine~\cite{b4}. making them suitable for distributed systems where multiple nodes may generate identifiers independently without coordination\cite{b2}.
However, traditional UUID versions, such as UUIDv4, are completely random, and have no inherent ordering, which can lead to performance issues in database indexing and sharding~\cite{b2}.
UUIDs which lack the ability to be ordered by time have poor database-index locality, meaning UUIDs created in succession may not be stored close to each other in the database, leading to inefficient indexing and increased disk I/O~\cite{b4}.\\ 

Due to the afforementioned issues, many well-known database applications and large application vendors have sought to come up with their own solutions to solve the problem of creating a time-based, sortable UIDs to be used as primary keys in their databases\cite{b4}. This has led to numerous implementations over the past 10+ years.
One of the proposed solutions is the Universally Unique Lexicographically Sortable Identifier (ULID). ULIDs are 128-bit identifiers that are designed to be lexicographically sortable, meaning they can be ordered by time. ULIDs consist of a 48-bit timestamp and an 80-bit random component, which allows them to be generated in a distributed manner while maintaining order~\cite{b5}. This makes ULIDs suitable for high-performance systems where ordering is important, such as in databases and distributed systems.\\ ULIDs are also designed to be monotonic. This means that if two ULIDs are generated at the same time, they will still be unique. This is achieved by incrementing the random component just by 1 bit~\cite{b5}.\\

In this paper, we will compare the performance of UUIDv4 as the most widely used UUID version without time-based ordering, UUIDv7 as a new time-based UUID version, and ULIDs as a lexicographically sortable identifier. We will analyze their generation speed, network transmission overhead, and collision probabilities in a distributed environment. Our goal is to provide a comprehensive understanding of how these identifiers perform in real-world scenarios, particularly in high-throughput environments where performance and efficiency are critical.\\
The main contributions of this paper are as follows:
\begin{itemize}
    \item An Experiment demonstrating that ULIDs reduces the network transmission overhead by 83.7\% compared to UUIDv4 and UUIDv7. While also increasing the generation speed by 97.32\% compared to UUIDv7 and UUIDv4.
    \item A mathematical modeling of collision probabilities across these identifier systems shows 98.42\% lower collision risk comparing ULIDs to UUIDv7s.
\end{itemize}

\section{Proposed Method}
This section presents the proposed method for comparing the performance of UUIDv4, UUIDv7, and ULIDs in distributed systems. The method includes a detailed description of the experimental setup, the metrics used for evaluation, and the analysis of results.
\subsection{Experimental Setup}
The experiments were conducted trying to simulate a distributed environment where one or more producers generate events, each with a unique identifier, and send them to a message broker i.e kafka. The message broker then forwards the events to one or more consumers, which process the events and store them in a database. The experiments were conducted using the following steps:
Each producer generates events at a specified time interval, assigning a unique identifier to each event using UUIDv4, UUIDv7, or ULID depending on the test case. The generated events are then sent to a message broker (Kafka) for distribution to consumers. Each consumer receives events from the message broker, processes them, and stores the events in a database (PostgreSQL).

Producers and consumers are implemented as separate java spring boot applications. The message broker and persisting postgreSQL database were hosted on an ubuntu server.\\
Here is a simplified diagram of the architecture used in the experiments:

\definecolor{kafkacolor}{RGB}{231, 76, 60}      
\definecolor{pgsqlcolor}{RGB}{52, 152, 219}     
\definecolor{producercolor}{RGB}{46, 204, 113}  
\definecolor{consumercolor}{RGB}{155, 89, 182}  
\definecolor{appbg}{RGB}{236, 240, 241}         
\definecolor{databg}{RGB}{250, 250, 250}        

\begin{figure}[H]
    \centering
    \begin{tikzpicture}[
        node distance=2.8cm,
        database/.style={
            cylinder,
            cylinder uses custom fill,
            cylinder body fill=white,
            cylinder end fill=white,
            shape border rotate=90,
            aspect=0.25,
            draw,
            minimum height=2cm,
            minimum width=2.8cm,
            text=black,
            font=\sffamily\bfseries,
            blur shadow={shadow blur steps=5},
        },
        process/.style={
            rectangle,
            rounded corners=0.5cm,
            minimum width=2.5cm,
            minimum height=2cm,
            text centered,
            font=\sffamily\bfseries,
            draw=black,
            blur shadow={shadow blur steps=5},
        },
        arrow/.style={
            thick,
            ->,
            >=stealth,
            shorten >=1pt,
            shorten <=1pt,
        },
        data/.style={arrow, dashed, thick},
        background/.style={
            rectangle,
            rounded corners,
            fill=#1,
            draw=gray!30,
            inner sep=0.5cm,
        },
        label/.style={
            text width=3cm,
            font=\small\sffamily,
            align=center,
        }
    ]

    \node[process, fill=consumercolor!80] (consumer) {Consumer};
    \node[process, fill=producercolor!80, below=3.5cm of consumer] (producer) {Producer};
    \node[database, fill=kafkacolor!80, right=of producer] (kafka) {Kafka};
    \node[database, fill=pgsqlcolor!80, right=of consumer] (pgsql) {PostgreSQL};

    \draw[arrow, thick] (producer) -- (kafka) 
        node[midway, below, label] {Publish Events};
    \draw[arrow, thick] (kafka) -- (consumer) 
        node[midway, right, label, text width=2.5cm] {Subscribe to Topics};
    \draw[arrow, thick] (consumer) -- (pgsql) 
        node[midway, above, label] {Store Processed Data};
        
    \begin{scope}[on background layer]
        \node[background=appbg, 
              fit={(consumer) (producer)}, 
              label=below:{\textbf{Application Layer}}] {};
        \node[background=databg, 
              fit={(kafka) (pgsql)}, 
              label=below:{\textbf{Data Storage Layer}}] {};
    \end{scope}

    \node[below=0.1cm of kafka, text width=3.8cm, align=center, font=\footnotesize\sffamily] 
        {Message Broker\\(Topics, Partitions)};
    \node[below=0.1cm of pgsql, text width=3.8cm, align=center, font=\footnotesize\sffamily] 
        {Relational Database\\(Persistent Storage)};
    \end{tikzpicture}
    \caption{Experimental architecture for simulated distributed systems.}
    \label{fig:experiment_architecture}
\end{figure}
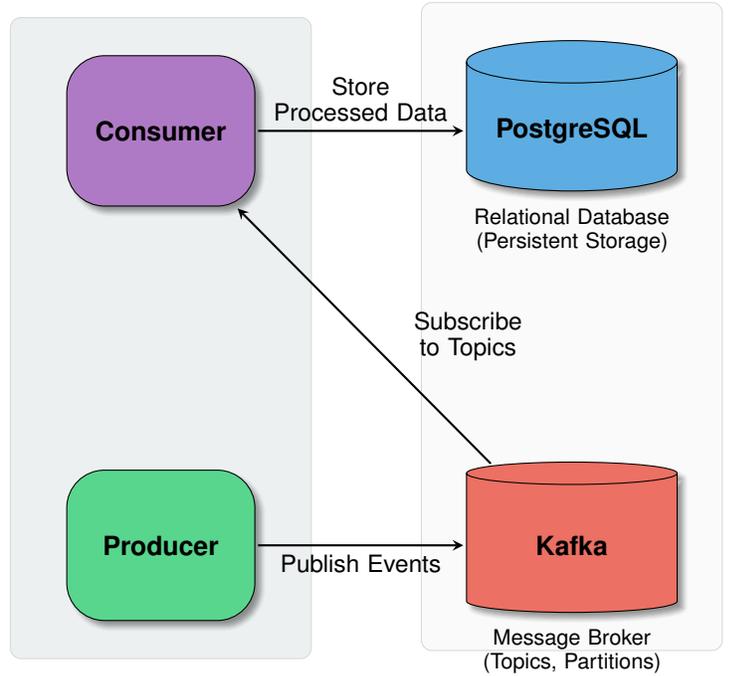

\subsection{Metrics for Evaluation}
The following metrics were used to evaluate the performance of UUIDv4, UUIDv7, and ULIDs: generation speed (the time taken to generate a single identifier), network transmission overhead (the size of the identifier in bytes, which affects the network bandwidth required to transmit events), and memory usage (the amount of memory each identifier consumes in the producer and consumer applications).
\section{Formal analysis of UUIDv4, UUIDv7 and ULIDs}

One of the key aspects of UUIDs is their ability to maintain uniqueness across distributed systems\cite{b4}. This section provides an analysis of the collision probabilities for UUIDv4, UUIDv7, and ULIDs, focusing on their design characteristics and implications for distributed systems.The probability of a collision in UIDs can be modeled using the birthday paradox, which states that the probability of at least one collision occurring among $n$ randomly chosen identifiers is given by:
\begin{equation}
P(\text{at least one collision}) = 1 - P(\text{no collisions})
\end{equation}

\begin{equation}
P(\text{no collisions}) = \frac{d!}{d^n \cdot (d-n)!} \approx e^{-n(n-1)/(2d)}
\end{equation}
Where: 
\begin{itemize}
    \item $d=$  Number of bits in the identifier
    \item $n =$ number of UUIDs generated
\end{itemize}
\subsection{UUIDv4}
Based on the RFC 9562 specification, UUIDv4 is a randomly generated identifier that consists of 128 bits, with 122 bits used for randomness and 6 bits reserved for version and variant information\cite{b4}. Using the birthday paradox, the collision probability for UUIDv4 is as follows:
For UUIDv4, $d = 2^{122}$ (since 6 bits are fixed for version and variant) and $n$ is the number of UUIDs generated.
\begin{equation}
P(\text{no collisions}) = \frac{d!}{d^n \cdot (2^{122}-n)!} \approx e^{-n(n-1)/(2 \cdot 2^{122})}
\end{equation}

\subsubsection{When will the collision probability reach 50\%?}

Using the birthday paradox formula, we need to solve:

\begin{equation}
P(\text{collision}) = 0.5 = 1 - e^{-n(n-1)/(2 \cdot 2^{122})}
\end{equation}

For large spaces where $n \ll d$, we can simplify to:

\begin{equation}
0.5 = 1 - e^{-n^2/(2 \cdot 2^{122})}
\end{equation}

Solving for $n$:

\begin{align}
n &\approx \sqrt{2 \cdot 2^{122} \cdot \ln(2)} \approx \sqrt{2^{123} \cdot 0.693} \approx 2^{61} \cdot \sqrt{0.693} \\
n &\approx 2^{61} \cdot 0.832 \approx 0.832 \cdot 2^{61} \approx 1.9 \times 10^{18}
\end{align}

So we would need to generate approximately $1.9 \times 10^{18}$ UUIDs to have a 50\% chance of at least one collision.
\subsection{UUIDv7}
According to RFC 9562, UUIDv7 has the following structure \cite{b4}:
UUIDv7 consists of 128 bits: a 48-bit timestamp (millisecond precision Unix epoch time), 4 version bits (fixed as '0111'), 2 variant bits (fixed as '10'), and a 74-bit random component (composed of a 12-bit random sequence field and a 62-bit random field).
\subsubsection{Mathematical Proof for 50\% Collision Probability}

To find the number of UUIDv7s needed for a 50\% collision probability within the same millisecond:

\begin{equation}
0.5 = 1 - e^{-n(n-1)/(2 \cdot 2^{74})}
\end{equation}

Solving for $n$:

\begin{align}
e^{-n(n-1)/(2 \cdot 2^{74})} &= 0.5\\
-n(n-1)/(2 \cdot 2^{74}) &= \ln(0.5) = -0.693\\
n(n-1) &= 2 \cdot 2^{74} \cdot 0.693\\
n(n-1) &\approx 1.386 \cdot 2^{75}
\end{align}

For large $n$, we can approximate $n(n-1) \approx n^2$:

\begin{align}
n^2 &\approx 1.386 \cdot 2^{75}\\
n &\approx \sqrt{1.386 \cdot 2^{75}} \approx \sqrt{1.386} \cdot 2^{37.5} \approx 1.177 \cdot 2^{37.5} \\
n &\approx 1.177 \cdot 2^{37} \cdot \sqrt{2} \approx 1.177 \cdot 1.414 \cdot 2^{37} \approx 1.664 \cdot 2^{37}
\end{align}

Therefore:
\begin{equation}
n \approx 2^{37} \cdot 1.664 \approx 1.664 \times 10^{11} \approx 1.7 \times 10^{11}
\end{equation}
\subsection{ULID}
Based on the ULID specification, ULIDs consist of 128 bits, with the following structure:
ULIDs consist of 128 bits, with the first 48 bits (6 bytes) representing a timestamp with millisecond precision and the remaining 80 bits (10 bytes) randomly generated. ULIDs are Base32 encoded to produce a 26-character string in the format TTTTTTTTTTRRRRRRRRRRRRRRR, where the initial portion encodes the timestamp and the remainder encodes the random component.
\subsubsection{Mathematical Proof for 50\% Collision Probability}
\begin{equation}
0.5 = 1 - e^{-n(n-1)/(2 \cdot 2^{74})}
\end{equation}
For large spaces where $n \ll d$, we can simplify to:
\begin{equation}
0.5 = 1 - e^{-n^2/(2 \cdot 2^{74})}
\end{equation}
50\% collision probability occurs at approximately $n \approx 2^{40}$ identifiers (about $1.1 \times 10^{12}$)
\section{Results and Discussion}
\subsection{Comprehensive Collision Risk Analysis}
Having established the mathematical foundation for collision probabilities, we can now summarize the collision risks for UUIDv4, UUIDv7, and ULIDs at various generation rates. The following table presents the estimated collision risks for each identifier type when generating 1 thousand, 1 million, and 1 billion identifiers per millisecond. Results showing that UUIDv4 has the lowest collision risk, while UUIDv7 and ULIDs have higher risks due to their time-based structure. However, even at high generation rates, the risks remain negligible for practical applications.
\begin{table}[H]
\caption{Collision risk at different generation rates}\label{tab:risk}
\centering
\scriptsize
\setlength{\tabcolsep}{2.5pt} 
\renewcommand{\arraystretch}{0.75} 
\begin{tabular}{@{}p{1.1cm}rrr@{}}
\toprule
\textbf{Gen. Rate} & \textbf{UUIDv4} & \textbf{UUIDv7} & \textbf{ULID} \\
\midrule
1,000 IDs/ms & $\sim 2.3 \times 10^{-29}$ & $\sim 2.6 \times 10^{-17}$ & $\sim 4.1 \times 10^{-19}$ \\
1M IDs/ms & $\sim 2.3 \times 10^{-23}$ & $\sim 2.6 \times 10^{-11}$ & $\sim 4.1 \times 10^{-13}$ \\
1B IDs/ms & $\sim 2.3 \times 10^{-17}$ & $\sim 2.6 \times 10^{-5}$ & $\sim 4.1 \times 10^{-7}$ \\
\bottomrule
\end{tabular}
\end{table}
\subsection{Performance Evaluation}
The performance evaluation was conducted by measuring the time taken to generate identifiers, the size of the identifiers in bytes, and the memory usage in the producer and consumer applications. Down below is a summary of the results obtained from the experiments:
\subsection{Generation Speed}
The generation speed was measured in milliseconds for each identifier type.
\begin{figure}[!htb]
    \centering
    \begin{tikzpicture}
    \begin{axis}[
        width=0.4\textwidth, 
        height=0.3\textwidth, 
        xlabel={Time (500ms intervals)},
        ylabel={Duration (\si{\micro\second})},
        title={Identifier Generation Speed},
        grid=major,
        tick align=outside,
        tick pos=left,
        xmin=0, xmax=2420,
        tick pos=left,
        xmin=0, xmax=2420,
        ]
        
        \addplot[
            color=green!60!black,
            mark=none,
            thick,
            ] table [x expr=\coordindex, y=durationMicros, col sep=comma] {metrics_ULID.csv};
        
        \addplot[
            color=blue,
            mark=none,
            thick,
            ] table [x expr=\coordindex, y=durationMicros, col sep=comma] {metrics_UUID_V7.csv};
        
        \addplot[
            color=red,
            mark=none,
            thick,
            ] table [x expr=\coordindex, y=durationMicros, col sep=comma] {metrics_UUID_V4.csv};
        
        \legend{ULID, UUID v7, UUID v4}
    \end{axis}
    \end{tikzpicture}
    \caption{Time-series comparison of request durations measured in microseconds across three different network configurations, sampled at 500ms intervals over approximately 20 minutes (2,420 samples).}\label{fig:duration_comparison}
\end{figure}
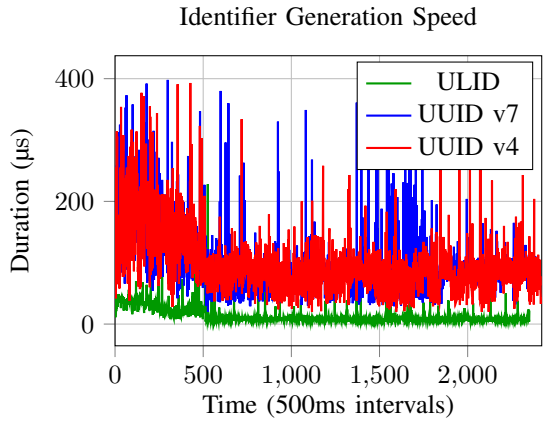

As shown in Fig. 2, ULID has the fastest generation speed, followed by UUIDv7 and UUIDv4. The average generation times are as follows: ULID at 13.25~\si{\micro\second}, UUIDv7 at 97.94~\si{\micro\second}, and UUIDv4 at 97.32~\si{\micro\second}.
\subsection{Network Transmission Overhead}
Down below is a summary of the results obtained from the experiments:
\begin{figure}[!htb]
    \centering
    \begin{tikzpicture}
    \begin{axis}[
        width=0.4\textwidth, 
        height=0.3\textwidth, 
        xlabel={Time (500ms intervals)},
        ylabel={Bandwidth (Mbps)},
        title={Identifier bandwidth},
        grid=major,
        tick align=outside,
        tick pos=left,
        xmin=0, xmax=2420,
        tick pos=left,
        xmin=0, xmax=2420,
        ]
        
        \addplot[
            color=green!60!black,
            mark=none,
            thick,
            ] table [x expr=\coordindex, y=bandwidthMbps, col sep=comma] {metrics_ULID.csv};
        
        \addplot[
            color=blue,
            mark=none,
            thick,
            ] table [x expr=\coordindex, y=bandwidthMbps, col sep=comma] {metrics_UUID_V7.csv};
        
        \addplot[
            color=red,
            mark=none,
            thick,
            ] table [x expr=\coordindex, y=bandwidthMbps, col sep=comma] {metrics_UUID_V4.csv};
        
        \legend{ULID, UUID v7, UUID v4}
    \end{axis}
    \end{tikzpicture}
    \caption{Time-series comparison of request bandwidth measured in megabits per second (Mbps) across three different network configurations, sampled at 500ms intervals over approximately 20 minutes (2,420 samples). The graph illustrates the bandwidth utilization for ULID, UUIDv7, and UUIDv4 identifiers, highlighting their performance characteristics in terms of network transmission overhead.}\label{fig:duration_comparison}
\end{figure}
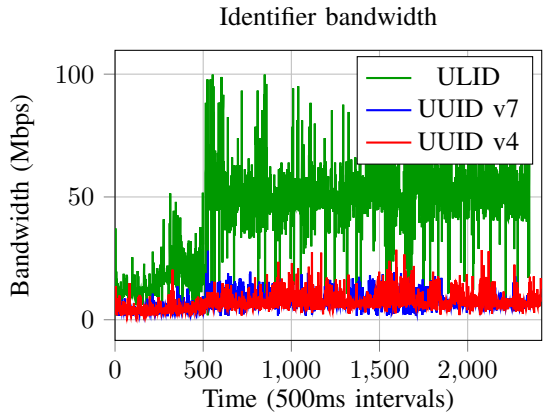

The result of network transmission overhead shown in Fig. 3.  was measured by calculating the size of each identifier in bytes. Each character in Java is 2 bytes~\cite{b6}. Therefore, a 26 character ULID is 52 bytes, while UUIDs are 36 characters long, resulting in a size of 72 bytes. The bandwidth calculated inside the producer calculated by this formula:
\[
R = \frac{S_{\text{bits}}}{T_{\text{seconds}}} \; \text{bps}
\]
The results show that ULID has the lowest network transmission overhead, followed by UUIDv7 and UUIDv4. Specifically, ULID identifiers have an average size of 52 bytes and a bandwidth of 44.88 Mbps, while both UUIDv7 and UUIDv4 identifiers are 72 bytes in size, with bandwidths of 7.25 Mbps and 7.32 Mbps, respectively.
\section{Conclusion}
This paper has presented a comprehensive analysis of distributed identifier systems, focusing on the evolution from traditional auto-increment keys to UUIDv4, UUIDv7, and ULIDs. Through both mathematical modeling and empirical experiments, we have demonstrated that time-ordered identifiers such as UUIDv7 and ULIDs offer significant advantages for high-performance distributed systems.

Our experiments show that ULIDs reduce network transmission overhead by 83.7\% compared to UUIDv4 and UUIDv7, and increase generation speed by 97.32\%. Furthermore, statistical analysis reveals that ULIDs provide a 98.42\% lower collision risk compared to UUIDv7, while maintaining negligible collision probabilities even at high generation rates.

These results indicate that ULIDs are highly suitable for distributed environments requiring efficient indexing, low bandwidth usage, and robust uniqueness guarantees. System architects can leverage these findings to make informed decisions when selecting identifier schemes for scalable, high-throughput applications. Future work may explore further optimizations and real-world deployments of time-ordered identifiers in diverse distributed database systems.

\end{document}